\documentstyle[eqsecnum,aps]{revtex}

\begin{document}
\draft \preprint{HEP/123-qed}
\title{Polaron Crossover in Molecular Solids}
\author{ Marco Zoli$^1$ and A. N. Das $^2$}
\address{$^1$ Istituto Nazionale Fisica della Materia, Universit\'a di
Camerino,\\ 62032, Italy. - marco.zoli@unicam.it}

\address{$^2$ Theoretical Condensed Matter Physics Division,
Saha Institute of Nuclear Physics, \\ 1/AF Bidhannagar, Calcutta
700064, India. - atin@cmp.saha.ernet.in}

\date{\today}

\maketitle

\begin{abstract}
An analytical variational method is applied to the molecular
Holstein Hamiltonian in which the dispersive features of the
dimension dependent phonon spectrum are taken into account by a
force constant approach. The crossover between a large and a small
size polaron is monitored, in one, two and three dimensions and
for different values of the adiabatic parameter, through the
behavior of the effective mass as a function of the
electron-phonon coupling. By increasing the strength of the
inter-molecular forces the crossover becomes smoother and occurs
at higher {\it e-ph} couplings. These effects are more evident in
three dimensions. We show that our Modified Lang-Firsov method
starts to capture the occurence of a polaron self-trapping
transition when the electron energies become of order of the
phonon energies. The self-trapping event persists in the fully
adiabatic regime. At the crossover we estimate polaron effective
masses of order $\sim 5 - 40$ times the bare band mass according
to dimensionality and value of the adiabatic parameter. Modified
Lang-Firsov polaron masses are substantially reduced in two and
three dimensions. There is no self-trapping in the antiadiabatic
regime.
\end{abstract} 

\narrowtext
\section*{I. Introduction}

There has been a growing interest towards polarons over the last
years also in view of the technological potential of polymers and
organic molecules \cite{hanoi,lu} in which polaronic properties
have been envisaged. Theoretical investigations on polarons
generally start from the Holstein Hamiltonian \cite{holst}
originally proposed for a  diatomic molecular chain along which
hopping of electrons, linearly coupled to the vibrational quanta,
takes place according to a tight-binding description. If the local
{\it e-ph} coupling is sufficiently strong the induced lattice
deformation may dress the electron and transform it into a
polaronic charge carrier \cite{landau,toyozawa}. The conditions
for polaron formation and its mobility properties may however
depend also on the adiabaticity ratio, on dimensionality, on
peculiarities and anharmonicities of the lattice structure
\cite{devreese,deraedt,kopida,fehske,pucci,tsiro,romero1,mello,ale,voulga}.
As the Holstein Hamiltonian can be identically applied to states
made of excitons \cite{rashba,mishchenko} and phonons it also
provides a useful tool in optical spectra analysis and transport
properties of organic materials \cite{hoffmann,nguyen} with large
scale applications.

While the physical properties of polaronic systems change
\cite{emin,alemott} according to the size of the polaronic
quasiparticle, a number of theoretical tools
\cite{jeckel,romero,ciuchi,barisic,trugman} has been applied to
clarify nature and width of the crossover between a large (with
respect to the lattice constant) polaron at weak {\it e-ph}
coupling and a small polaron at strong coupling for a given value
of the adiabatic parameter. Provided that a phase transition is
ruled out in the Holstein Hamiltonian with dispersive optical
phonons \cite{lowen} being the ground state energy analytic in the
{\it e-ph} coupling, such a crossover may still appear as a smooth
transition in the antiadiabatic regime or rather as a sudden (but
continuous) event in the adiabatic regime \cite{eagles}. While the
narrowing of the polaron band signals the onset of the crossover
it is certainly from the behavior of the effective mass that the
self-trapping event can be accurately located in the intermediate
region of {\it e-ph} couplings \cite{zoli1}. Although precisely in
the latter region perturbative studies traditionally reveal their
shortcomings an analytical method, the Modified Lang-Firsov (MLF)
transformation, \cite{atin1} has been developed to overcome the
limitations of the standard Lang-Firsov (LF) \cite{lang} approach
on which strong coupling perturbation theory is based. As an
enhancement of the polaron mass should be accompanied by a
reduction in the polaron size, the electron-phonon correlation
function offers an independent tool to analyse the crossover
through the measure of the spread of the lattice deformation.
Since the notion of self-trapping transition has often assumed
different meanings in the literature we emphasize that our view of
a {\it self-trapped polaron} is not that of a localized and
immobile object but, rather, of a small quasiparticle whose
different ground state properties have undergone a transition,
driven by the {\it e-ph} coupling, at distinct although closely
related points in the polaron parameter space  \cite{bonca}. Among
these properties we study in this paper, using the MLF
transformation \cite{dasil,daschoudhury}, the polaron mass and the
correlation function as obtained from a Holstein Hamiltonian in
which the {\it dispersion} of the optical phonon branches is fully
accounted for in any dimensionality. Besides depicting a model
more appropriate to physical systems, dispersive phonons represent
a relevant feature of the Holstein model itself as previously
shown by one of us \cite{zoli2}. The role of the intermolecular
forces in the crossover of the MLF polarons at different
dimensionalities is a main focus of our investigation. The
generalities of the model are given in Section II while the
results are presented in Section III both for the polaron mass and
for the static correlation function. Section IV contains some
final remarks.

\section*{II. Modified Lang-Firsov Phonon Basis for the Holstein model}

We consider the dimension dependent Holstein Hamiltonian
consisting of one electron hopping term, an interaction which
couples the electronic density and the ionic displacements at a
given site and dispersive harmonic optical phonons as:

\begin{eqnarray}
H &=& - t \sum_{<i j>} c_i^{\dag} c_{j}
    +  g  \sum_i n_i (b_i^{\dag} + b_i)
        + \sum_{\bf q} \omega_{\bf q} b_{\bf q}^{\dag} b_{\bf q}
\end{eqnarray}

the first sum is over $z$ nearest neighbors, $c_i^{\dag}$ and
$c_i$ are the real space electron creation and annihilation
operators, $n_i~( = c_i^{\dag} c_i)$ is the number operator,
$b_i^{\dag}$ and $b_i$ are the phonons creation and annihilation
operators. $b_{\bf q}^{\dag}$ is the Fourier transform of
$b_i^{\dag}$ and $\omega_{\bf q}$ is the frequency of the phonon
with vector momentum ${\bf q}$.

The standard practice in dealing with the Hamiltonian (1) is to
apply the LF transformation where a phonon basis of fixed
displacements (at the electron residing site) is used. Such a
choice of phonon basis diagonalizes the Hamiltonian in absence of
hopping. The hopping term is then treated as a perturbation.
However the LF approach under simple approximations, $e.g.$ within
zero-phonon averaging or zeroth order of perturbation, cannot
describe the retardation between the electron and the lattice
deformations produced by the electron. This retardation induces a
spread in the size of the polaron and becomes very important for
weak and intermediate $e$-ph coupling. The MLF phonon basis, where
the displacements of the oscillators at different sites around an
electron are treated variationally, can describe the retardation
and a large to small polaron crossover even within simple
approximations \cite{dasil,daschoudhury}. Recently the convergence
of the perturbation series within the LF and the MLF methods has
been studied in a two-site Holstein model for the ground state
\cite{atin1} as well as for the first excited state \cite{JCD3}.
It was found that: (i) within the MLF method the perturbation
corrections are much smaller than those corresponding to the LF
method in the range from weak to intermediate $e$-ph coupling,
(ii) the convergence of the perturbation series within the MLF is
also much better in that range, (iii) in the strong coupling limit
the MLF phonon basis reduces to the LF basis and the LF
pertubation method works very well in this limit. The above
studies have clearly pointed out that the MLF perturbation method
works much better than the LF method when the entire range of the
$e$-ph coupling is considered.

The MLF perturbation method has also been applied to a many-site
Holstein model with dispersionless phonons in 1D  and the
supremacy of the MLF method over the LF method in predicting the
ground state energy and dispersion of the polaron has been
observed \cite{atin2}. For the present case of dispersive phonon
we apply the MLF transformation to the dimension dependent
Hamiltonian (1):

\begin{equation}
\tilde{H} = e^R H e^{-R}
\end{equation}

where

\begin{eqnarray}
R & = & \sum_{\bf q} \lambda_{\bf q} n_{\bf q}  (b_{-{\bf
q}}^{\dag} - b_{\bf q}), \nonumber
\\
n_{\bf q} &= & \frac{1}{\sqrt N}\sum_i n_i e^{-i{\bf q} \cdot R_i}
    = \frac{1}{\sqrt N}\sum_{\bf k} c_{\bf k+q}^{\dag} c_{\bf k}
\end{eqnarray}

and $\lambda_{\bf q}$s are the variational parameters which
represent the shifts of the equilibrium positions of the
oscillators (quantized ion vibrations) with momentum ${\bf q}$.
For conventional Lang-Firsov transformation $\lambda_{\bf
q}=g/\omega_{\bf q}$. The MLF transformed Hamiltonian  for a
single electron case is obtained as

\begin{eqnarray}
\tilde{H} &=& - \epsilon_p \sum_i n_i
    - t_p \sum_{{i j}} ~c_i^{\dag} c_{j}~\nonumber \\
      & &\times {\rm{exp}}[\frac{1}{\sqrt N}
        \sum_{\bf q} \lambda_{\bf q} b_{\bf q}^{\dag}(e^{i{\bf q}
        \cdot R_i}-e^{i{\bf q} \cdot R_j})] \nonumber \\
       & &\times {\rm{exp}}[-\frac{1}{\sqrt N}
        \sum_{\bf q} \lambda_{\bf q} b_{\bf q} (e^{-i{\bf q}
        \cdot R_i}-e^{-i{\bf q} \cdot R_j})] \nonumber \\
       & +& \sum_{\bf q} \omega_{\bf q} b_{\bf q}^{\dag} b_{\bf q}
    + \sum_{\bf q} (g-\lambda_{\bf q} \omega_{\bf q}) n_{\bf q} (b_{-{\bf q}}^{\dag} + b_{\bf q})
\end{eqnarray}

where

\begin{equation}
\epsilon_p = \frac{1}{N}~\sum_{\bf q} (2 g - \lambda_{\bf q}
\omega_{\bf q}) \lambda_{\bf q}
\end{equation}

is the polaron self-energy and

\begin{equation}
t_p = t~{\rm{exp}} \bigl[ -\frac{1}{N}
\sum_{\bf q} \lambda_{\bf q}^2 (1-\frac{\gamma_{\bf q}}{z}) \bigr]
\end{equation}

is the polaronic hopping. The coordination number $z$ is twice the
system dimensionality.

\begin{eqnarray}
\gamma_{\bf q} = \sum_j^{\prime} e^{i {\bf q} R_{ij}}
=  2 \sum_{i=x,y,z}cosq_i   \nonumber
\end{eqnarray}
where i and j are nearest neighbor sites. As unperturbed
Hamiltonian we choose $H_0$ as

\begin{equation}
H_0= -\epsilon_p \sum_i n_i + \sum_{\bf q} \omega_{\bf q} b_{\bf
q}^{\dag}b_{\bf q}
\end{equation}

The remaining part of the Hamiltonian ($\tilde H-H_0$) in the MLF
basis is considered as the perturbation part. The energy
eigenstates of $H_0$ are given by

\begin{equation}
|\phi_i, \{n_{\bf q}\}\rangle = c_i^{\dag} |0 \rangle_e
 |n_{{\bf q}_1}, n_{{\bf q}_2}, n_{{\bf q}_3}, ...\rangle_{ph}
\end{equation}

where, $i$ is the electron site and $n_{{\bf q}_1}, n_{{\bf q}_2},
n_{{\bf q}_3}$ are the phonon occupation numbers in the phonon
momentum states ${\bf q}_1, {\bf q}_2, {\bf q}_3$, respectively.
The lowest energy eigenstate of the unperturbed Hamiltonian has no
phonon excitations, $i.e.~ n_{\bf q}=0$ for all ${\bf q}$. The
ground state has an energy $E_0^0=- \epsilon_p$ and is $N$-fold
degenerate, where $N$ is the number of sites in the system. The
perturbation lifts the degeneracy and to first order in $t$ the
ground state energy of the 3D- polaron with momentum ${\bf k}$ is
given by

\begin{equation}
E_0({\bf k})= -\epsilon_p - t_p \gamma _{\bf k}
\end{equation}

and the corresponding eigenstate is $|{\bf k},{n_{\bf q}=0}
\rangle = \frac{1}{\sqrt{N}} \sum_i e^{i{\bf k} \cdot R_i}
c_i^{\dag} |0 \rangle_e |0\rangle_{ph}$.

The second order correction to the ground-state energy of the
polaron with momentum ${\bf k}$ is given by

\begin{equation}
E_0^{(2)}({\bf k})= \sum_{\bf k'}~ \sum_{\{n_{\bf q}\}}
\frac{1}{\sum_{\bf q} n_{\bf q} \omega_{\bf q}} |\langle \{n_{\bf
q}\},{\bf k'}|\tilde {H} - H_0|{\bf k},\{0\}\rangle |^2
\nonumber \\
\end{equation}

It is evident that the second order correction has contributions
from intermediate states having all possible phonon numbers,
$i.e.$ each $n_{\bf q}$ in Eq. (10) takes values from zero to
infinity with the condition that $n_{TOT}=\sum_{\bf q} n_{\bf q}
\ge 1$.

By minimizing the zone center ground state energy we get the
variational parameters $\lambda_{\bf q}$:

\begin{equation}
\lambda_{\bf q} = \frac{g}{\omega_{\bf q} + z t_p (1-
\frac{\gamma_{\bf q}}{z})}
\end{equation}

and, by such a choice of $\lambda_{\bf q}$, the one phonon matrix
element between the ground state $|{\bf k=0},\{n_{\bf
q}=0\}\rangle$ and the first excited state

\begin{eqnarray}
\langle 1_& &{\bf q},{\bf k'}|\tilde {H}- H_0|{\bf
k=0},\{0\}\rangle = \delta_{{\bf k'},-{\bf q}}
\frac{1}{\sqrt{N}}\times \nonumber \\ & &[-z t_p \lambda_{\bf q}
\bigl(1-\frac{\gamma_{\bf q}}{z} \bigr) +
  (g- \lambda_{\bf q} \omega_{\bf q})] \nonumber \\
\end{eqnarray}

vanishes. Then, the one phonon excitation process yields no
contribution to the second order correction for the MLF ground
state energy. The $\lambda_{\bf q}$'s appropriate to the 1D, 2D
and 3D systems are easily obtained by (11).

The static correlation function involving the electron charge at
{\it i-th} site and the lattice deformation at the {\it i+n} site
are given by

\begin{equation}
\chi_n = \langle \psi_G|c_i^{\dagger}c_i (b_{i+n}^{\dagger}
+b_{i+n})|\psi_G \rangle /2 \bar{g} \langle n_i\rangle
\end{equation}

where $\bar{g}={N^{-1}} \sum_{\bf q} (g/\omega_{\bf q})$ and
$|\psi_G\rangle$ denotes the ground state for the polaron with
momentum ${\bf k}$=0. The denominator in Eq. (13) is used to
normalize the correlation function with respect to its on-site
value in the strong coupling limit.
$n_i$ is the electron number
operator and $\langle n_i\rangle$= 1/N for the 1-electron system.
While $c_i$ and $b_i$ are the bare electron and phonon
annihilation operators in the undisplaced oscillator basis
respectively, the corresponding operators in the MLF basis are the
annihilation operators for the polaron and that of the phonon in
the variationally displaced oscillator basis. The correlation
function is calculated in the MLF basis within zero phonon
averaging.

\section*{III. Polaron Mass and Correlation Functions}

Previous investigations have pointed out that the Holstein model
with a dispersionless spectrum ($\omega_1=0$) or with weak
intermolecular forces ($\omega_1 \ll \omega_0$) would predict {\it
larger polaron bandwidths in lower dimensionality} against
physical expectations \cite{zoli2}. Moreover, as pointed out by
Holstein in his original papers \cite{holst}, dispersionless
phonons would lead to an unphysical {\it divergent site jump
probability} for the polaronic quasiparticle \cite{yama}. Hence,
intermolecular forces are a {\it key ingredient} of the Holstein
model.

Numerical analysis \cite{zoli2} has shown that the bandwidths
$\Delta E_d$ grow faster versus the intermolecular energy
$\omega_1$ in higher dimensionality $d$ thus providing a criterion
to fix the minimum $\omega_1$ which ensures the validity of the
Holstein model. Imposing the inequalities criterion $\Delta E_{3D}
\ge \Delta E_{2D} \ge \Delta E_{1D}$ we set the {\it threshold
value} $\bar \omega_1$ which turns out to be a function of the
breathing mode energy $\omega_0$ and of the $d-$independent {\it
e-ph} coupling $g_0=g/\sqrt{d}$ ($g$ scales $\propto \sqrt{d}$):
thus, at intermediate $g_0$ ($\simeq 1 - 1.5$, in units of
$\omega_0$) $\bar \omega_1$ is $\simeq \omega_0/2$, while at
strong $g_0$ ($\ge 2$) $\bar \omega_1$ should be at least $\simeq
2 \omega_0/3$ in order to ensure the correct bandwidths trend. On
the other hand, the intermolecular energies encounter the upper
bound $\omega_1 < \omega_0$ given by the value of the coupling
energy between the two atoms in the basic unit of the molecular
solid. Moreover, too large $\omega_1$ may invalidate strong
coupling perturbative treatments of the Holstein model for three
dimensional systems in fully adiabatic regimes \cite{zoli3}.

With this caveat we study the polaron mass both in the Lang-Firsov
and in the Modified Lang-Firsov scheme taking a lattice model in
which first neighbors molecular sites interact through a force
constants pair potential. Then, the $d-$ dependent optical phonon
spectrum is given by

\begin{eqnarray}
\omega^2_{1D}(q)=& &\, {{\alpha + \gamma } \over M} + {1 \over M}
\sqrt { \alpha^2 + 2 \alpha  \gamma cosq + \gamma^2} \nonumber \\
\omega^2_{2D}({\bf q})=& &\, {{\alpha + 2 \gamma } \over M} +
 {1 \over M} \sqrt { \alpha^2 +
2 \alpha  \gamma g({\bf q}) +  \gamma ^2 (2 + h({\bf q}))}
\nonumber \\ \omega^2_{3D}({\bf q})=& &\, {{\alpha + 3 \gamma }
\over M} +
 {1 \over M} \sqrt { \alpha^2 +
2 \alpha  \gamma j({\bf q}) + \gamma ^2 (3 + l({\bf q}))}
\nonumber
\\ g({\bf q})=& &\,cosq_x + cosq_y \nonumber
\\  h({\bf q})=& &\, 2cos(q_x -
q_y) \nonumber \\ j({\bf q})=& &\,cosq_x + cosq_y + cosq_z
\nonumber
\\l({\bf q})=& &\, 2cos(q_x - q_y) + 2cos(q_x - q_z) + 2cos(q_y -
q_z)) \nonumber
\\
\end{eqnarray}

where the intra-molecular force constant $\alpha$ and the
inter-molecular first neighbors force constant $\gamma$ are
related to  $\omega_0$ and $\omega_1$ by $\omega_0^2=\,2\alpha/M$
and $\omega_1^2=\,\gamma/M$ respectively. $M$ is the reduced
molecular mass. In terms of $\omega_0$, the dimensionless
parameter $zt/\omega_0$ defines the adiabatic ($zt/\omega_0 > 1$)
and the antiadiabatic ($zt/\omega_0 < 1$) regime.

Second order perturbative theory introduces the polaron mass $m^*$
dependence on the hopping integral $t$, hence on the adiabatic
parameter, which would be absent in the first order Lang-Firsov
theory. Generally, $m^*$ can vary with $t/\omega_0$ in two ways:
$m^*$ becomes lighter either by increasing $\omega_0$ at fixed $t$
or, by increasing $t$ at fixed $\omega_0$. As the mass variation
due to $\omega_0$ is much stronger than that due to $t$, for a
given adiabatic parameter, we may get different mass values
according to the absolute values of $\omega_0$ and $t$. However,
for sufficiently strong {\it e-ph} couplings which make the
perturbative method applicable, the LF mass changes only slightly
with $t$ and second order corrections are small unless the
intramolecular phonon energies are low ($\omega_0 < 50meV$)
\cite{zoli3}. Hereafter we set $\omega_0=100meV$ and select the
adiabatic parameter by tuning $t$.

In Figs.1, we plot the ratio of the one dimensional polaron mass
to the bare band mass against the {\it e-ph} coupling calculated
both in the Lang-Firsov scheme  and in the Modified Lang-Firsov
expression.

An intermediate regime $2t=\,\omega_0$ is assumed in Fig.1(a)
while the intermolecular energy spans a range of weak to strong
values. The striking different behavior between the LF and the MLF
mass occurs for intermediate $g$ while at very strong couplings
the MLF plots converge, as expected, towards the LF predictions.
The LF method overestimates the polaron mass for $g \in [\sim 1 -
2]$ (according to the value of $\omega_1$) and mostly, it does not
capture the rapid mass increase found instead in the MLF
description. Note that, around the crossover, the MLF polaron mass
is of order ten times the bare band mass in the case
$\omega_1=\,60meV$. Large intermolecular energies enhance the
phonon spectrum thus reducing the effective masses in both
figures. In the MLF method, large $\omega_1$ tend also to smooth
the mass behavior in the crossover region.

Going to a fully adiabatic regime (see Fig.1(b)) the discrepancies
between LF and MLF plots are even more pronounced and the range of
{\it e-ph} couplings in which the two methods converge shrinks
considerably. There is scarce renormalization in the MLF curves up
to the crossover which is clearly signalled by a sudden {\it
although continuous} mass enhancement whose abruptness is
significantly smoothed for the largest values of intermolecular
energies. In the antiadiabatic case shown in Fig.1(c), the picture
changes drastically and we recover a nearly coincident mass
behavior in the LF and MLF methods throughout the whole range of
couplings. The convergence is favoured at large $\omega_1$. As
above mentioned the LF plots show a strong resemblance in going
from Fig.1(a) to Fig.1(c): infact, the LF method slightly depends
on the hopping integral in 1D systems with large intramolecular
energy. The results we have displayed so far induce to reconsider
the concept of {\it self-trapping} traditionally indicating an
abrupt, but continuous, transition between an infinite size states
at weak {\it e-ph} couplings and a finite (small) size polaron at
strong {\it e-ph} couplings. According to the adiabatic polaron
theory \cite{emholst,tsiro} there is no self-trapping event in one
dimension as the polaron solution is always the ground state of
the system. Instead, in higher dimensionality a minimum coupling
strength is required to form finite size polarons, hence
self-trapped polarons can exist at couplings larger than that
minimum. As a shrinking of the polaron size yields a weight
increase, the polaron mass behavior is accepted to be the most
reliable indicator of the self-trapping transition. The latter
appears to us as a crossover essentially dependent on the degree
of adiabaticity of the system and crucially shaped by the internal
structure of the phonon cloud which we have modelled by tuning the
intermolecular forces. We are then led to relocate the self
trapping event in the parameter space of 1D systems admitting that
also finite size polarons can self-trap if a sudden change in
their effective mass occurs for some values of the {\it e-ph}
couplings in some portions of the adiabatic regime. As
fluctuations in the lattice distortions around the electron site
are included in our variational wave function discontinuities in
the polaron mass should not appear at the onset of the transition
\cite{shore}. Mathematically we select the crossover points
through the simultaneous occurence of a maximum in the first
logarithmic derivative and a zero in the second logarithmic
derivative of the MLF polaron mass with respect to the coupling
parameter: such inflection points, corresponding to the points of
most rapid increase for $m^*$, are reported on in Figures 2, where
the mass ratios are plotted for a wide choice of antiadiabatic to
adiabatic regimes and a sizeable value of $\omega_1$ both in one,
two and three dimensions.

Some well known features of the antiadiabatic polaron landscape
are confirmed by our analytical variational model in all
dimensionalities: i) antiadiabatic polarons are generally heavier
than adiabatic ones although, at very strong couplings, the mass
values converge at the Lang-Firsov results and ii) there is no
self-trapping in the fully antiadiabatic regime as the electron
and the dragged phonon cloud form a compact unit, a small polaron,
also at intermediate {\it e-ph} couplings. Then, the mass increase
is smooth in the antiadiabatic regime. Instead, in the more
controversial \cite{zheng} antiadiabatic to adiabatic transition
region we start to detect the signatures of the crossover which
persist in the fully adiabatic regime and form a line of
self-trapping events whose features however change considerably
versus dimensionality. In 1D (Fig.2(a)), the crossover occurs for
$g$ values between $\sim 1.8 - 2.3$ and the corresponding self
trapped masses are of order $\sim 5 - 50$ times the bare band mass
thus suggesting that relatively light small polarons can exist in
1D molecular solids with high phonon spectrum. The self-trapped
mass values grow versus $g$ by increasing the degree of
adiabaticity and the incipience of the self-trapping line is set
at the intermediate value $2t/\omega_0=\,1$. We note that these
findings are in good qualitative and quantitative agreement with
refined variational results supporting the existence of
self-trapped polarons also in 1D. Although in the deep adiabatic
regime we find a quasi step-like increase, the 1D polaron mass is
a continuous and derivable function of the {\it e-ph} coupling
\cite{romero}.

The two dimensional lattice introduces some significant novelties
in the MLF mass behavior as shown in Fig.2(b): i) at a given {\it
e-ph} coupling and adiabaticity ratio, the 2D mass is lighter than
the 1D mass and the 2D LF limit is attained at a value which is
roughly one order of magnitude smaller than in 1D; ii) the
crossover region is shifted upwards along the $g$ axis with the
self trapping events taking place in the range, $g$ $\sim 2.2 -
2.6$ and the corresponding masses are of order $\sim 5 - 10$ times
the bare band mass; iii) the curve connecting the self trapping
points is parabolic with an extended descending branch starting at
the intermediate value $4t/\omega_0=\,1$; iv) in the deep
adiabatic regime, the lattice dimensionality smoothens the mass
increase versus $g$. The latter effect is even more evident in 3D,
see Fig.2(c), as there are no signs of abrupt mass increase even
for the largest values of the adiabatic parameter. At the
crossover, 3D masses are of order $\sim 5 - 10$ times the bare
band mass with the self trapping points lying in the range, $g$
$\sim 2.5 - 2.9$. At large couplings the {\it effective mass over
bare band mass} ratio becomes independent of the $t$ value and
converges towards the LF value. In this region (and for the choice
$\omega_1=\,60meV$) the 3D Lang-Firsov mass is one order of
magnitude smaller than the 2D mass. As the coordination number
grows versus dimensionality, large intermolecular forces are more
effective in hardening the 3D phonon spectrum thus leading to
lighter 3D polaron masses than 2D ones.

In Figures 3 we plot the correlation functions $\chi_0$, $\chi_1$
and $\chi_2$ in 1D (a) and 2D (b) respectively, as obtained by
(13) for the adiabatic regime $zt/\omega_0=\,2$ with
$\omega_0=100meV$. Two values, $\omega_1=\,40meV$ and
$\omega_1=\,80meV$, have been chosen to point out the role of the
intermolecular forces in the transition between a large polaron at
weak couplings and a small polaron at strong couplings. For
sufficiently strong $g$ values the LF limit is obtained, {\it
i.e.} $\chi_0$ becomes 1 while $\chi_1$ and $\chi_2$ become zero
implying that the resulting polaron is an on site small polaron.
The small to large polaron cross-over is manifested by a strong
reduction of $\chi_0$ alongwith an enhancement in the values of
$\chi_1$ and $\chi_2$. By increasing $\omega_1$, the crossover is
slightly smoothed and shifted upwards along the $g$ axis.
Accordingly, $\chi_1$ and (to a lesser extent) $\chi_2$ acquire
some weight throughout a larger portion of {\it e-ph} coupling
values. As a main feature we note that the crossovers indicated by
the correlation functions of the one dimensional system, for the
two selected cases, occur at $g/\omega_0 \sim\,2$ and $\sim\,2.35$
respectively. These values match the corresponding crossover
points extracted by the polaron mass slopes. In two dimensions,
the self-trapping transition takes place at larger (than in 1D)
$g$ values and non local {\it e-ph} correlations persist in the
adiabatic polaron up to $g/\omega_0 \sim\,3$. The crossover is
generally smooth and the softening effect of the intermolecular
forces is more pronounced than in the one dimensional system.

\section*{IV. Conclusions}

We have developed a variational analytical method to study the
Holstein polaron problem versus dimensionality in the entire range
of (anti)adiabatic parameters characterizing the molecular system.
The essential role of the phonon dispersion in the Holstein model
has been accounted for including the intermolecular interactions
by means of a force constant approach. Unlike the traditional
Lang-Firsov scheme the Modified Lang-Firsov method permits to
describe the fact that, in the intermediate and adiabatic regimes,
the lattice deformation does not follow instantaneously the
electron motion thus leading to a spreading in the quasiparticle
size. Under these circumstances we have examined the behavior of
the polaron mass as a function of the strength of the {\it e-ph}
coupling and critically analysed the occurence of the
self-trapping event signalling a shrinking of the polaron size in
the real space. This crossover has been also monitored through the
computation of the static {\it e-ph} correlation functions which
provide a complementary tool corroborating our conclusions.
Varying the adiabatic parameter and selecting the points of most
rapid increase for the effective mass we have found a set of
self-trapping points originating, in 1D and 2D, in the
intermediate regime ($zt/\omega_0=\,1$) and continuing in the
fully adiabatic regime. In 3D, the self-trapping events occur at
$zt/\omega_0 >1$. While, in one dimension, the curve connecting
the inflection points in the adiabatic regime is a monotonic
growing function of the {\it e-ph} coupling, in two and three
dimensions we find distinctive parabola-like curves whose minima
(of order $\sim 5$ times the bare band mass) are located at larger
$g$ in higher $d$. Hence small polaron formation is favoured in
low $d$ whereas very large {\it e-ph} couplings are required to
shrink the size of adiabatic polarons in 3D. As intermolecular
forces play a stronger role in more closely packed structures,
lattice dimensionality is expected to shape the polaron behavior.
Infact, our results show that the crossover from large to small
polarons is, in 2D and even more in 3D, smoother than in the case
of the 1D adiabatic polaron at a fixed value of intermolecular
energy. Pointed out the quantitative differences in the polaron
mass according to the dimensionality one should however notice a
qualitative similarity in all dimensions regarding the occurence
of the self-trapping event. Finally we observe that, although
polaron masses become generally lighter in higher $d$, also in 1D
the {\it effective mass over bare band mass} ratio is $\sim 5$ at
the crossover when phonons and electrons compete on the energy
scale. Small polarons having mobility properties may be therefore
expected in low dimensional molecular systems with sufficiently
strong intermolecular forces.

\begin{figure} \vspace*{10truecm} \caption{Ratio of the
one dimensional polaron mass to the bare band mass versus {\it
e-ph} coupling according to the Lang-Firsov and the Modified
Lang-Firsov methods. The adiabatic parameter is set at: (a) the
intermediate value, $2t/\omega_0=\,1$; (b) a fully adiabatic
regime, $2t/\omega_0=\,2$; (c) an antiadiabatic regime,
$2t/\omega_0=\,0.25$. $\omega_0=\,100meV$ and $\omega_1$ (in units
$meV$) are the {\it intramolecular} and {\it intermolecular}
energies of the diatomic molecular chain respectively.}
\end{figure}

\begin{figure}
\vspace*{15truecm} \caption{Ratio of the Modified Lang-Firsov
polaron mass to the bare band mass versus {\it e-ph} coupling in
(a) 1D, (b) 2D and (c) 3D. A set of twelve $zt/\omega_0$ values
ranging from the antiadiabatic to the adiabatic regime is
considered. From left to right: $zt/\omega_0=\,0.25, 0.5, 0.75,
1.0, 1.25, 1.5, 1.75, 2.0, 2.25, 2.5, 2.75, 3.0$.
$\omega_0=\,100meV$. The diamonds mark the occurence of the
self-trapping event.}
\end{figure}

\begin{figure}
\vspace*{10truecm} \caption{(a) One dimensional and (b) two
dimensional static correlation functions versus {\it e-ph}
coupling in the adiabatic regime, $zt/\omega_0=\,2$.
$\omega_0=\,100meV$. Two values of intermolecular energies,
$\omega_1=\,40meV$ and $\omega_1=\,80meV$, have been taken.}
\end{figure}

\section*{Acknowledgements}

This work is part of the Joint Research Project (Ph-T 4) under the
Indo-Italian Programme of Co-operation in Science and Technology
2002-2004.  A.N.D. acknowledges hospitality at the Physics
Department, University of Camerino. M.Z. acknowledges hospitality
at the Saha Institute of Nuclear Physics (Calcutta) and thanks
Mr.B. Ram for his kind help during his stay in the SINP Guest House.

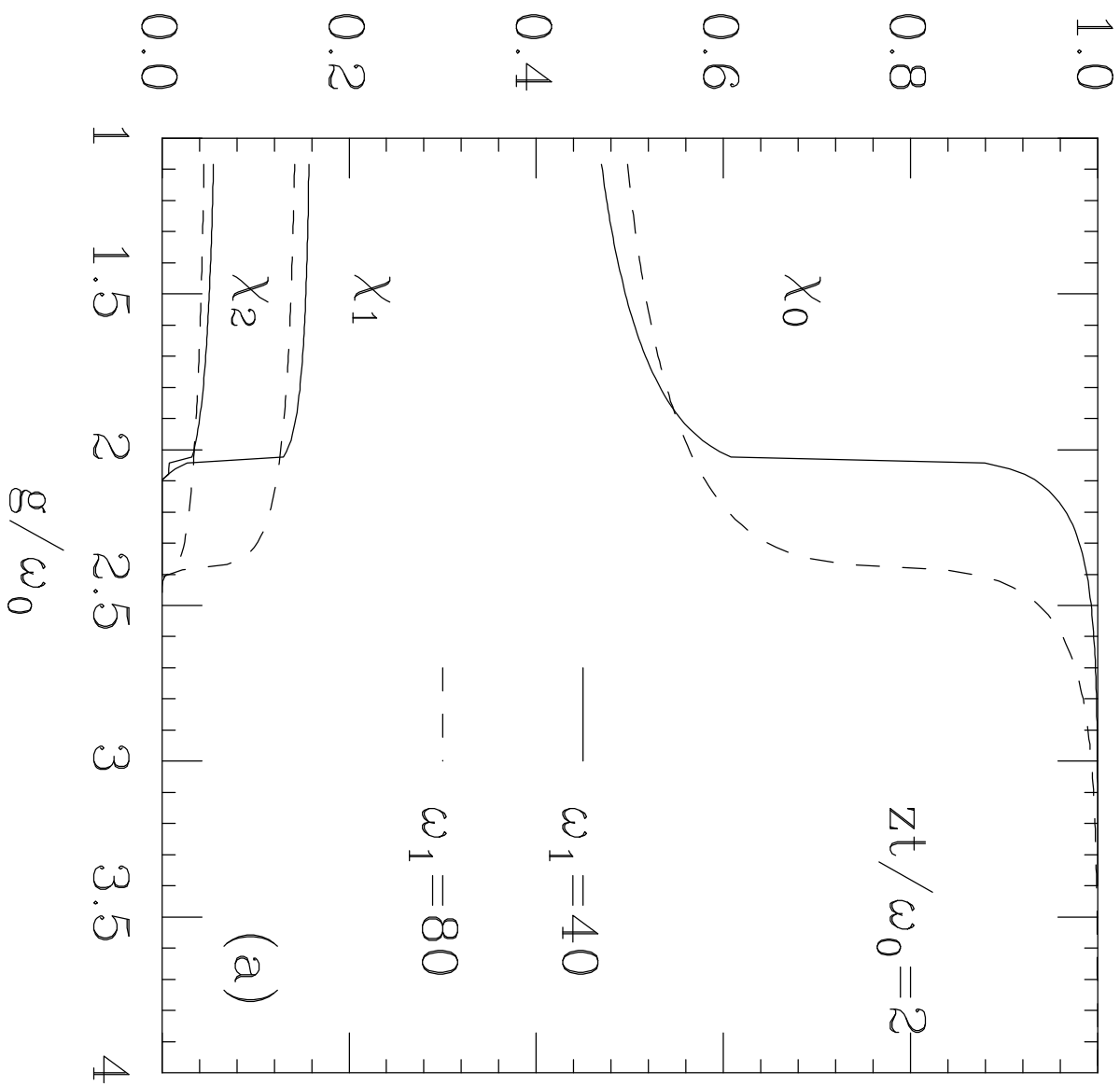
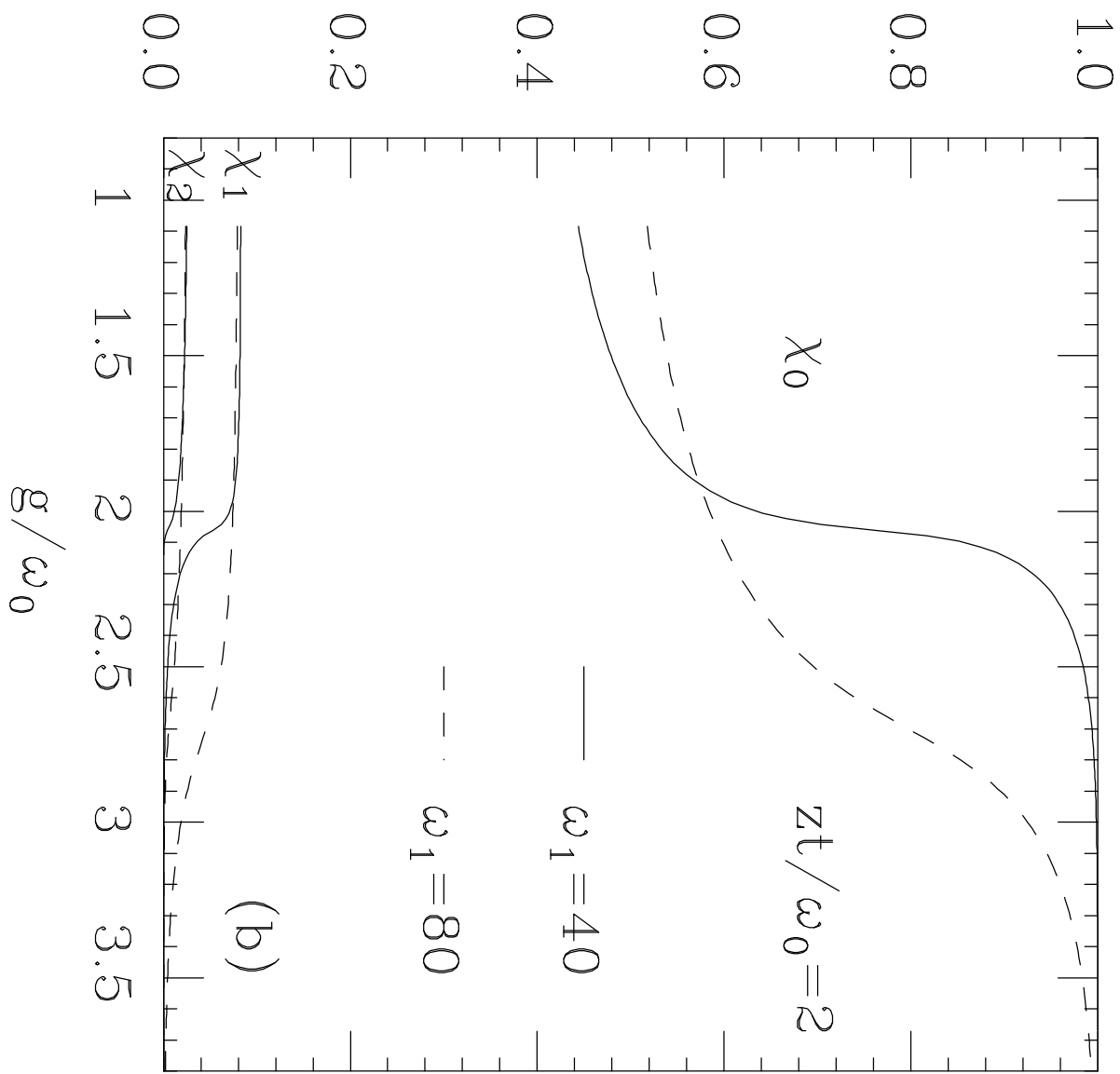
\end{document}